\documentclass[manuscript]{aastex631}
\usepackage{amsmath,amstext}
\usepackage{graphicx}
\usepackage{natbib}
\usepackage{amssymb}
\usepackage{makecell}
\usepackage{multirow}
\usepackage{threeparttable}
\newcommand{\Alf}{Alfv$\acute{\rm e}$n}

\received{}
\revised{}
\accepted{}

\shortauthors{Shi et al.}
\begin{document}

\title{Excitation of Multi-periodic Kink Motions in Solar Flare Loops: Possible Application to Quasi-periodic Pulsations}

\correspondingauthor{Mijie Shi}
\email{shimijie@sdu.edu.cn}

\author{Mijie Shi}
\affiliation{Shandong Key Laboratory of Optical Astronomy and Solar-Terrestrial Environment, School of Space Science and Physics, Institute of Space Sciences, Shandong University, Weihai, Shandong, 264209, China}

%\author{Tom Van Doorsselaere}
%\affiliation{Centre for mathematical Plasma Astrophysics, Department of Mathematics, KU Leuven, B-3001 Leuven, Belgium}

\author{Bo Li}
\affiliation{Shandong Key Laboratory of Optical Astronomy and Solar-Terrestrial Environment, School of Space Science and Physics, Institute of Space Sciences, Shandong University, Weihai, Shandong, 264209, China}

\author{Shao-Xia Chen}
\affiliation{Shandong Key Laboratory of Optical Astronomy and Solar-Terrestrial Environment, School of Space Science and Physics, Institute of Space Sciences, Shandong University, Weihai, Shandong, 264209, China}

\author{Mingzhe Guo}
\affiliation{Shandong Key Laboratory of Optical Astronomy and Solar-Terrestrial Environment, School of Space Science and Physics, Institute of Space Sciences, Shandong University, Weihai, Shandong, 264209, China}
\affiliation{Centre for mathematical Plasma Astrophysics, Department of Mathematics, KU Leuven, 3001 Leuven, Belgium}

\author{Shengju Yuan}
\affiliation{Institute of Frontier and Interdisciplinary Science, Shandong University, Qingdao, Shandong, 266237, China}



\begin{abstract}
Magnetohydrodynamic (MHD) waves are often invoked to interpret 
	quasi-periodic pulsations (QPPs) in solar flares.
We study the response of a straight flare loop to a kink-like velocity perturbation
	using three-dimensional MHD simulations 
	and forward model the microwave emissions using the fast gyrosynchrotron code.
Kink motions with two periodicities are simultaneously generated,
	with the long-period component ($P_{\rm L}=57$~s) being attributed to the radial fundamental kink mode 
	and the short-period component ($P_{\rm S}=5.8$~s) to the first leaky kink mode.
Forward modeling results show that the two-periodic oscillations 
	are detectable in the microwave intensities for some lines of sight.
Increasing the beam size to $(1^{\prime\prime})^2$ does not wipe out the microwave oscillations. 
We propose that the first leaky kink mode
	is a promising candidate mechanism to account for short-period QPPs.
Radio telescopes with high spatial resolutions can help distinguish
	between this new mechanism with such customary interpretations as sausage modes.

\end{abstract}




\section{Introduction} 
\label{S-Introduction}
Quasi-periodic pulsations (QPPs) are oscillatory intensity variations commonly observed 
	in solar flare emissions across a broad range of passbands
	\citep[e.g.,][]{2010ApJ...723...25T,2010SoPh..267..329K,2011ApJ...740...90V,2015A&A...574A..53K,2021ApJ...921..179L}.
The typical periods of QPPs range from a fraction of a second to several minutes.
A variety of candidate mechanisms have been proposed to explain QPPs
	\citep[see reviews, e.g.,][]{2009SSRv..149..119N,2016SoPh..291.3143V,2018SSRv..214...45M,2020STP.....6a...3K,2021SSRv..217...66Z}.
However, the physical mechanisms responsible for generating QPPs remain uncertain.

Rapid QPPs, namely those QPPs with periodicities on the order of seconds,
	are customarily attributed to
	such mechanisms as oscillatory reconnection \citep[e.g.,][]{1991ApJ...371L..41C,1992ApJ...399..159H} 
	or sausage modes \citep[e.g.,][]{1983Natur.305..688R,2003A&A...412L...7N}.
Oscillatory reconnection can occur at a two-dimensional X-point 
	as a consequence of, say, the impingement by velocity pulses \citep{2009A&A...493..227M}.
This mechanism was shown to operate in the hot corona
	with the oscillatory periods independent of the initial velocity pulse \citep{2022ApJ...933..142K}.
Sausage modes can cause periodic compression and rarefaction and thus modulate
	the microwave emissions of flare loops \citep[e.g.,][]{
		2012ApJ...748..140M,2014ApJ...785...86R,
		2015SoPh..290.1173K,2021ApJ...921L..17G,
		2022MNRAS.516.2292K,2022ApJ...937L..25S}.
Candidate sausage modes have been reported in observations of QPPs in various passbands
    (e.g., 
    \citeauthor{2005A&A...439..727M}~\citeyear{2005A&A...439..727M};
    \citeauthor{2010SoPh..263..163Z}~\citeyear{2010SoPh..263..163Z};
    \citeauthor{2012ApJ...755..113S}~\citeyear{2012ApJ...755..113S};
    \citeauthor{2015A&A...574A..53K}~\citeyear{2015A&A...574A..53K};
    \citeauthor{2016ApJ...823L..16T}~\citeyear{2016ApJ...823L..16T};
    see the recent review by \citeauthor{2020SSRv..216..136L}~\citeyear{2020SSRv..216..136L}).
    
QPPs with multiple periodicities are often observed \citep{2010ApJ...723...25T}.
Magnetohydrodynamic (MHD) waves are accepted to play an important role for the formation of these multi-periodic signals. 
\cite{2009A&A...493..259I} interpreted their multi-periodic event as the result 
	of kink-mode-triggered magnetic reconnection.
Multiple periodic QPPs can also be attributed to the superposition of different modes,
	for example, fast and slow sausage modes \citep{2011ApJ...740...90V},
	axial fundamental kink mode and its axial overtones \citep{2013SoPh..284..559K},
	kink mode and sausage mode \citep{2015A&A...574A..53K}.
In these studies, the long-period (several tens of seconds) signals are often attributed to kink or slow sausage modes,
	whereas the short-period (several seconds) signals to fast sausage modes.  

In this work, we simulate kink motions in flare loops and forward model their microwave signatures,
	examining the possibility that both short- and long-period kink motions
	can be excited simultaneously.
We present the MHD setup and results in Section~\ref{S-MHD},
	moving on to describe the forward modeling setup and results in Section~\ref{S-GS}.
Section~\ref{S-summary} summarizes this study.

\section{MHD Simulation}
\label{S-MHD}
\subsection{Numerical Setup}
We model flare loops as field aligned, $z$-directed, axially uniform cylinders. 
%The equilibrium magnetic field is $z$-directed, 
%	and all equilibrium quantities are $z$-independent.
The equilibrium density distribution is prescribed by
\begin{eqnarray}
	\label{eq_def_rhoEQ}
\rho = \rho_{\rm e}+(\rho_{\rm i} - \rho_{\rm e})f(r),
\end{eqnarray}
	where $[\rho_{\rm i},\rho_{\rm e}]$=$[4\times 10^{10},1\times 10^{9}]m_{\rm p}~{\rm cm}^{-3}$
	represent the internal and external mass densities.
	In addition, $m_{\rm p}$ is the proton mass.
By `internal' (subscript i) and `external' (subscript e), we refer to the equilibrium quantities 
	at the loop axis and far from the loop, respectively.
A continuous profile $f(r) ={\rm exp}[-(r/R)^{\alpha}]$
	is used to connect $\rho_{\rm i}$ and $\rho_{\rm e}$,
	with $r=\sqrt{x^2+y^2}$, $\alpha=15$,
	and the nominal loop radius $R=3$~Mm.
The temperature distribution follows the same spatial dependence as the density,
	with $[T_{\rm i},T_{\rm e}]=[10,2]$~MK.
The magnetic field $\boldsymbol{B}$ is $z$-directed,
	its magnitude varying from $B_{\rm i}=60$~G to $B_{\rm e}=80$~G
	to maintain transverse force balance.
The pertinent Alfv\'en speeds are $[v_{\rm Ai},v_{\rm Ae}]=[654,5510]~\rm{km~s^{-1}}$.
The loop length is $L_0=30$~Mm.
The specification of our equilibrium is largely in accordance with typical measurements \citep[e.g.,][]{2016ApJ...823L..16T}.
%The corresponding sound speeds are  $c_{si}=526~\rm{km/s}$ and $c_{se}=235~\rm{km/s}$, respectively.
Figure~\ref{3D_view} shows the initial density distributions
	in the $z=L_0/2$ and $x=0$ cuts.
The equilibrium is perturbed by a kink-like initial velocity perturbation in the $x$-direction
\begin{eqnarray}
\label{eq2}
v_x(x,y,z;t=0)=v_0{\rm exp}\left(-\frac{r^2}{2\sigma^2}\right){\rm{sin}}\left(\frac{\pi z}{L_0}\right),
\end{eqnarray} 
where $v_0=20{\rm~km~s^{-1}}$ is the amplitude,
and $\sigma=R$ characterizes the spatial extent
   \footnote{The effect of $\sigma$ is also worth examining.
   The results are collected in Appendix~\ref{sec_AppB}
       to streamline the main text.}.

The set of ideal MHD equations is evolved using the finite-volume code PLUTO \citep{2007ApJS..170..228M}.
We use the piecewise parabolic method for spatial reconstruction,
	and the second-order Runge–Kutta method for time-stepping.
The HLLD approximate Riemann solver is used for computing inter-cell fluxes, 
	and a hyperbolic divergence cleaning method
	is used to keep the magnetic field divergence-free.
The simulation domain is $[-15,15]~\rm{Mm}\times[-15,15]~\rm{Mm}\times[0,30]~\rm{Mm}$.
A uniform grids with cell numbers of $[N_x,N_y,N_z] = [1000,1000,100]$ is adopted.
We use the outflow boundary condition for all primitive variables
	at the lateral boundaries ($x$ and $y$).
For the top and bottom boundaries ($z$),
	the transverse velocities ($v_x$ and $v_y$) are fixed at zero.
The density, pressure, and $B_z$ are fixed at their initial values.
The other variables are set as outflow.
Our numerical results vary little when, say, a finer grid or a larger domain 
	is experimented with. 
In particular, no spurious reflection is discernible at the lateral boundaries.

\subsection{Numerical Results}
\begin{figure}    %%%%%%%%%%%%%%%%%% FIGURE 2
\begin{center}
	\includegraphics[width=0.8\textwidth,clip=]{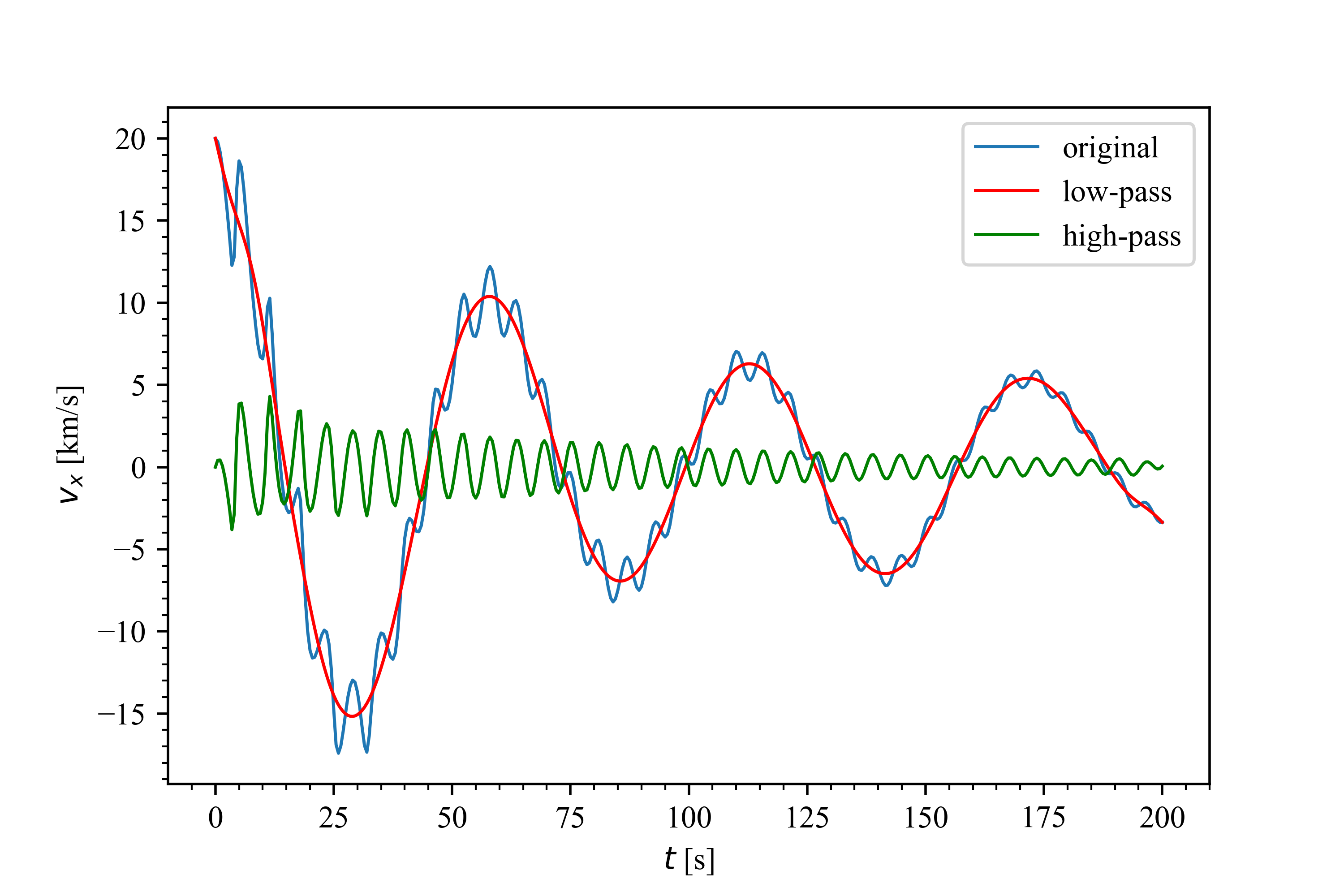}
	\caption{Temporal evolution of $v_x$ sampled at $(x,y,z)=(0,0,L_0/2)$ (blue),
		     together with its low-pass (red) and high-pass (green) components.}	
	\label{vx_apex}
\end{center}
\end{figure}

Figure~\ref{vx_apex} displays the temporal evolution of $v_x$ sampled 
	at the loop center $(x,y,z)=(0,0,L_0/2)$.
One sees that two periodicities co-exist.
We decompose the time sequence into a long-period (the red curve)
	and a short-period (green) component.
This decomposition is performed using the 
	low-(high-) pass filter of the {\it filtfilt} function in the {\it SciPy} package,
	with the threshold period chosen to be 10~s.
The periods of the two components are
	$P_{\rm L} = 57$~s and $P_{\rm S} = 5.8$~s, respectively.
Both components are seen to experience temporal damping.	

\begin{figure}    %%%%%%%%%%%%%%%%%% FIGURE 3
\begin{center}
	\includegraphics[width=0.8\textwidth,clip=]{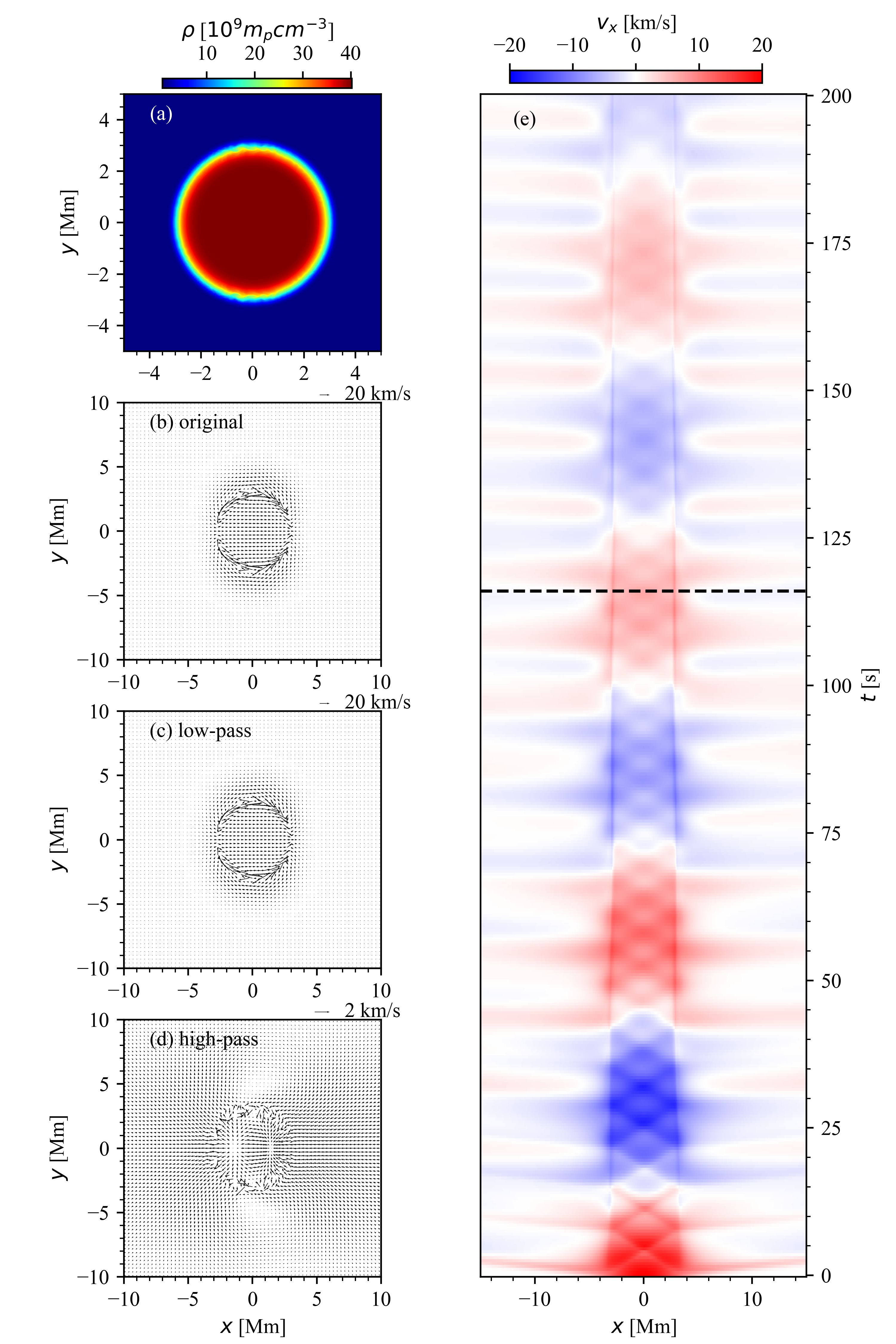}
	\caption{(a) Density distribution at the loop apex. (b) Velocity field along with its (c) long-period and (d) short-period components at the loop apex.
			(e) Temporal evolution of $v_x(x,y=0,z=L_0/2)$.
			The black dashed line marks the instant (116 s) where the figures in the left column are produced.
		An animated version of this
			figure is available that has the same layout as the static figure, and runs from 0–200s.}
	\label{vx_decompose}
	\end{center}
\end{figure}

The snapshot at $t=116$~s of 
the density at the loop apex is displayed in Figure~\ref{vx_decompose}(a).
Some rolled-up vortices can be readily seen at the loop boundary close
    to the $y$-axis, suggesting the development of
    the Kelvin-Helmholtz instability as a result of
    localized shearing motions
    \citep[e.g.,][]
    {2008ApJ...687L.115T,2010ApJ...712..875S,2015ApJ...809...72A}.
Figure~\ref{vx_decompose}(b) shows 
the velocity field at the loop apex, 
	together with its low-pass (Figure~\ref{vx_decompose}(c)) and high-pass (Figure~\ref{vx_decompose}(d)) components.
Figure~\ref{vx_decompose}(e) further displays the temporal evolution of $v_x(x,y=0,z=L_0/2;t)$.
The multi-periodic oscillation is clearly shown in the loop interior.
The period of $P_{\rm L} = 57$~s is consistent with that of the radial fundamental kink mode.
The low-pass component of the velocity field (Figure~\ref{vx_decompose}(c) and the related animation)
	is typical of a radial fundamental kink mode as well \citep[e.g.,][]{2014ApJ...788....9G}.
Therefore, the long-period oscillation is identified
    as the radial fundamental kink mode.
This identification is further corroborated by a comparison with 
    independent theoretical expectations from an eigenvalue problem perspective
    in Appendix~\ref{app_A1}, where we show that
    the associated temporal damping is attributable to
    resonant absorption \citep{2011SSRv..158..289G}.

The short-period component ($P_{\rm S} = 5.8$~s) shows up as ridge-like structures
	crossing each other in the loop interior shown in Figure~\ref{vx_decompose}(e).
The slope of each ridge matches the fast speed in the loop interior,
	indicating a fast wave nature of these ridges.
The high-pass component of the velocity field (Figure~\ref{vx_decompose}(d) and the related animation)
	shows some rapid variation of a dipole-like velocity field at the loop interior.
These features are consistent with the first leaky kink mode,
	i.e., the extension to the leaky regime of the first radial overtone
with the damping attributable to lateral leakage 
	\citep[e.g.,][]{1982SoPh...75....3S,1986SoPh..103..277C,2003SoPh..217...95C}.
The short-period oscillation associated with a first leaky kink mode is of particular interest
    as it has not been invoked for interpreting QPPs to our knowledge. 
Our numerical results show that the first leaky kink mode can be generated along with 
    the radial fundamental kink mode and this scenario is likely to happen in flare loops.
On this aspect we stress that the associated velocity field has not been demonstrated
    in initial value problem studies
    (see Appendix~\ref{app_A2} for further discussions).

\section{Forward Modeling the Gyrosynchrotron Emission}
\label{S-GS}
\subsection{Method}
\begin{figure}    %%%%%%%%%%%%%%%%%% FIGURE 1
	\begin{center}
		\includegraphics[width=0.8\textwidth,clip=]{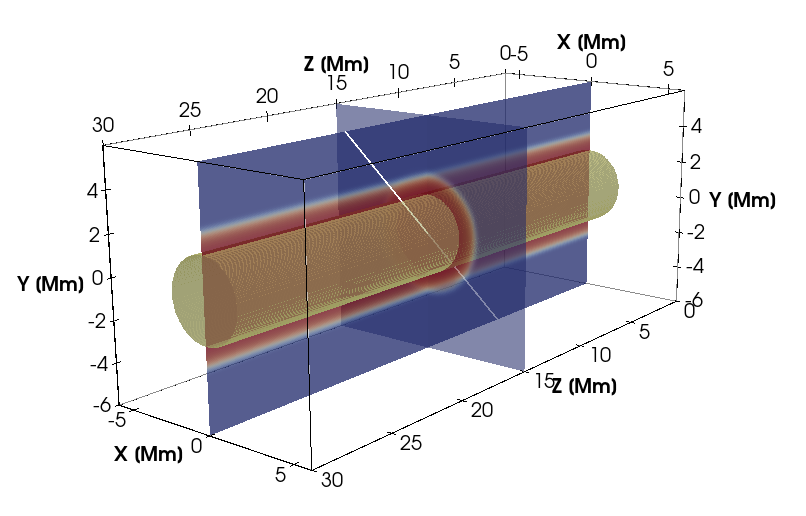}
		\caption{Initial density distributions in the $z=L_0/2$ and $x=0$ cuts. 
			     The yellow volume shows the region where the non-thermal electrons occupy at $t=0$.
				 The white line marks the line of sight (LoS) for the case $x_0=0$.}
		\label{3D_view}
	\end{center}
\end{figure}

We compute gyrosynchrotron (GS) emissions using the fast GS code \citep{2011ApJ...742...87K,2010ApJ...721.1127F}.
The fast GS code computes the local values of the absorption coefficient and emissivity,
    thereby accounting for inhomogeneous sources by integrating the radiative transfer equation.
We assume that non-thermal electrons occupy a time-varying volume that initially corresponds to 
	$r\le 2{~\rm Mm}$ (the yellow volume in Figure~\ref{3D_view}).
This assumption is based on the consideration that the non-thermal electrons
	determined by some acceleration mechanism may fill only part of the loop, 
	similar to the model used in \cite{2022MNRAS.516.2292K}.
This volume moves back and forth due to the kink motions.
We trace this volume using a passive scaler as the simulation goes on.
Within this volume,
	we further assume that the number density ($N_b$) of the non-thermal electrons is proportional to the thermal one ($N_e$),
    and specifically takes the form $N_b = 0.005N_e$.
The spectral index of the non-thermal electrons is $\delta=3$,
    with the energy ranging from $0.1$ to $10$~MeV.
The pitch angle distribution of the non-thermal electrons is
    taken to be isotropic.
We assume that any line of sight (LoS) is located in the $y-z$ plane,
	making an angle of $45^\circ$  with both the $y$- and $z$-axis.
The LoS intersects the apex plane $z=L_0/2$ at $(x_0,y_0)$
	and threads a beam with a cross-sectional area of 60~km$\times$60~km.
We select three beams with different $x_0$ (i.e., $x_0=0,\pm1$~Mm)
	by adopting a fixed $y_0=0$.
The white line in Figure~\ref{3D_view} illustrates the LoS for the case $x_0=0~\rm{Mm}$.

\subsection{Results}
\begin{figure}    %%%%%%%%%%%%%%%%%% FIGURE 4
	\begin{center}
		\includegraphics[width=0.8\textwidth,clip=]{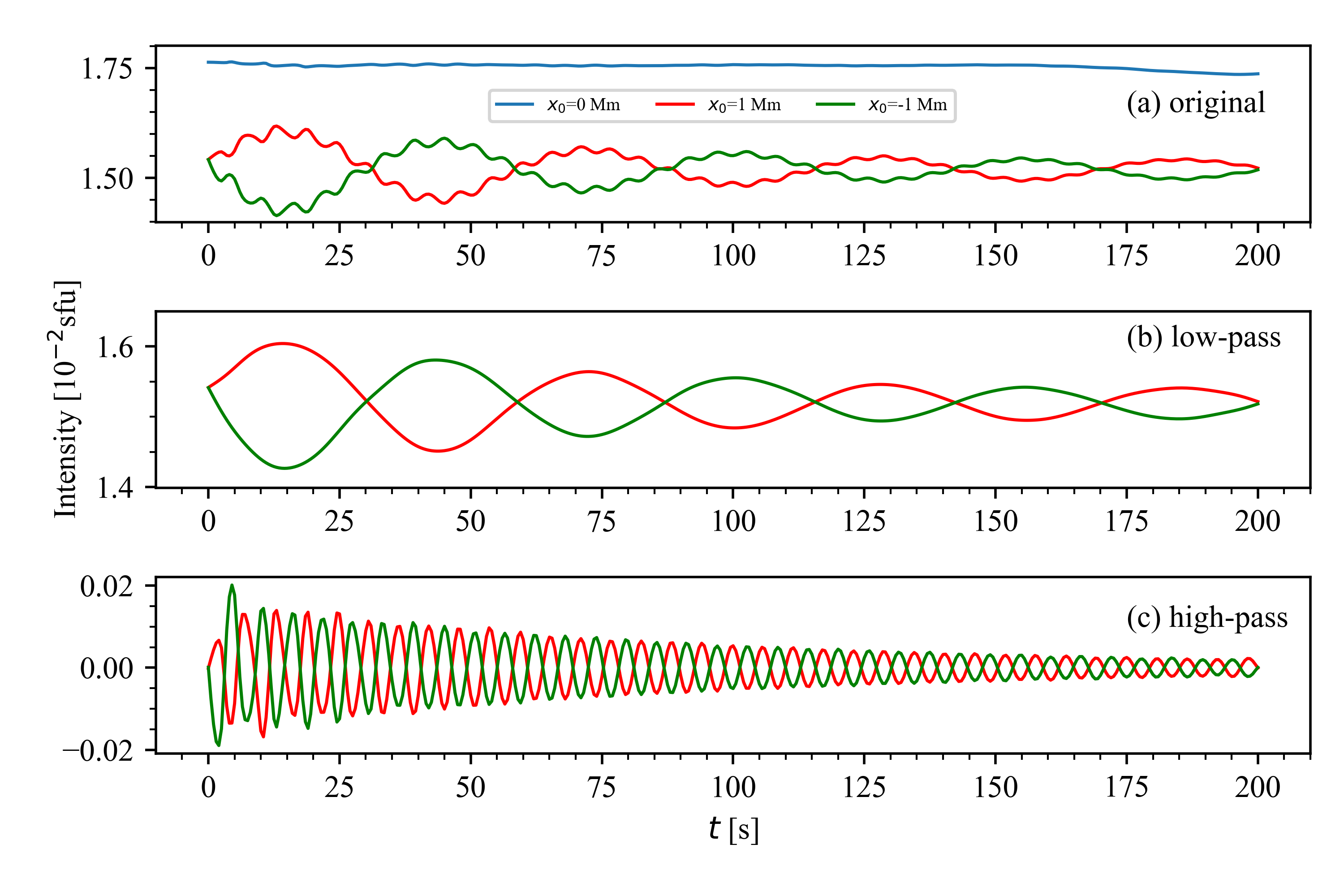}	
		\caption{(a) Intensity variations of the 17 GHz emission for three LoS beams.
				(b) Low-pass and (c) high-pass components of the intensity variations for $x_0=\pm1~\rm{Mm}$.}
		\label{int_17G}
	\end{center}
\end{figure}

We forward model the GS emission 
	and examine the microwave signature of the kink motions,
	taking the 17~GHz emission as an example.
Figure~\ref{int_17G}(a) shows the intensity variations for the three beams
	($x_0=0,\pm1$~Mm).
One sees that the intensity variation is negligible for the   
	beam that passes through the loop axis ($x_0=0$).
However, the intensity shows obvious variations with two different periodicities 
	for the rest of the beams ($x_0=\pm1~\rm{Mm}$).
We decompose the intensity variations into a long-period (Figure~\ref{int_17G}(b))
	and a short-period (Figure~\ref{int_17G}(c)) component for the cases $x_0=\pm1~\rm{Mm}$.
The periods of the two components are consistent with $P_{\rm L}$ and $P_{\rm S}$, 
	meaning that the two periodicities of the microwave intensity are
	the manifestations of the simultaneously excited radial fundamental kink mode and the first leaky kink mode.
We also find that the intensity variations are anti-correlated 
	between the cases $x_0=1~\rm{Mm}$ and $x_0=-1~\rm{Mm}$.
This behavior is caused by the asymmetric variations between the left and right sides of the loop
	for the two modes.
	
\begin{figure}    %%%%%%%%%%%%%%%%%% FIGURE 5
	\begin{center}
		\includegraphics[width=0.8\textwidth,clip=]{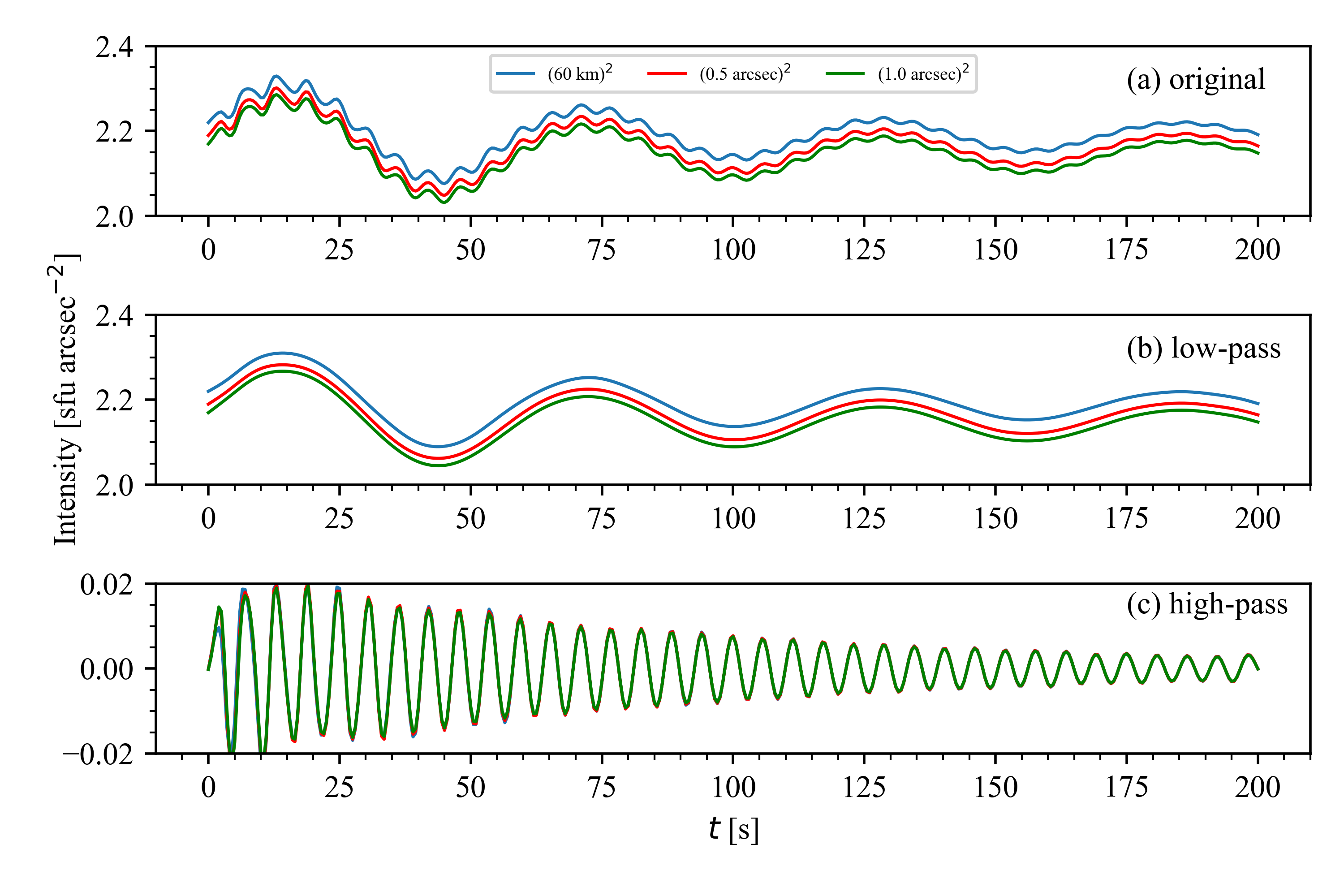}
		\caption{(a) Intensity variations of the 17 GHz emission for different beam sizes for the case $x_0=1~\rm{Mm}$.
				  (b) Low-pass and (c) high-pass components of (a).}
		\label{int_low_res}
	\end{center}
\end{figure}

Figure~\ref{int_low_res}(a) shows the intensity variations for the case $x_0=1~\rm{Mm}$
	when different beam sizes are adopted.
The intensity of a larger beam is achieved by adding the intensity of all single beams in the
	corresponding area projected onto the plane of sky.
Figures~\ref{int_low_res}(b) and \ref{int_low_res}(c)
	display the low-pass and high-pass component.
One sees that increasing the beam size to as large as $(1^{\prime\prime})^2$ does not
	change the two-periodicity behavior of the microwave intensities.
For the short-period component,
	the intensity variations are almost the same for different beam sizes.	

The kink motions with two periodicities are 
	likely to be detectable using modern radio telescopes.
The Mingantu Spectral Radioheliograph (MUSER, \cite{2021FrASS...8...20Y}) observes the Sun with a time cadence of 0.2 s
	and a spatial resolution of $1.3^{\prime\prime}$ at 15~GHz.
The Atacama Large Millimeter/submillimeter Array (ALMA, \cite{2016SSRv..200....1W}) can achieve unprecedented high resolutions
	at 85~GHz.

\section{Summary}
\label{S-summary}
We examined the response of a straight flare loop
	to a kink-like initial velocity perturbation using 3D MHD simulations.
We found that kink motions with two periodicities 
	are simultaneously generated in the loop interior.
The long-period component ($P_{\rm L}=57$~s) is attributed to the radial fundamental kink mode,
	and the short-period component ($P_{\rm S}=5.8$~s) to the first leaky kink mode.
We then examined the modulation of the microwave intensity by the kink motions
	via forward modeling the GS emissions.
We found that the two-periodic signals are detectable in the 17~GHz microwave emission
	at some LoS directions.
Increasing the beam size to as large as $(1^{\prime\prime})^2$ does not wipe out these oscillatory signals.

Leaky kink modes are a promising candidate mechanism to
	account for short-period QPPs in flare loops.
\cite{2015A&A...574A..53K} detected multi-periodic QPPs
	and interpreted the long-period component (100 s) as a radial fundamental kink mode
	and the short-period component (15 s) as a sausage mode.	
We argue that the short-period component in their observations can be 
	alternatively interpreted as the first leaky kink mode.
However, we cannot distinguish between the two mechanisms based purely on the oscillation period,
	because the timescales of the first leaky kink modes and sausage modes are close.
Radio telescopes with high spatial resolutions would be very helpful to identify the first leaky kink modes.
For sausage modes, the intensity variations are expected to be in-phase across the entire loop,
	whereas for the first leaky kink modes the intensity variations 
	are anti-correlated between the left and right sides of the loop.

\begin{acknowledgments}
We thank the referee for constructive comments.	
This work is supported by the National Natural Science Foundation of China (41904150, 12273019, 41974200, 11761141002, 12203030).
We gratefully acknowledge ISSI-BJ for supporting the international team “Magnetohydrodynamic wavetrains as a tool for probing the solar corona ”.
\end{acknowledgments}




\appendix
\section{Comparison Between 3D Time-dependent Results and Theoretical Expectations}
\label{app_A}
The temporal evolution of $v_x$ at three different locations
    in the apex plane ($z=L_0/2$)
    is displayed by the blue curves in the left column of 
    Figure~\ref{vx_parameter}.
Two periodicities can be readily seen in any time sequence,
    which is therefore decomposed 
	into a long-period (the red curves) and a short-period (green) component.
We then fit both components using an exponentially damping sinusoid
    $A_0 \sin(2\pi t/P+\phi_0){\rm exp}(-t/\tau)$,
    printing the best-fit periods ($P_{\rm L, S})$ and
    damping times ($\tau_{\rm L, S}$) on each plot. 
The best-fit curves are additionally displayed by
    the black dotted lines.

\subsection{the Long-period Component}
\label{app_A1}
The radial fundamental kink mode is well known to be resonantly absorbed in the 
    \Alf\ continuum in a radially continuous equilibrium with
    our ordering of the characteristic speeds
    \citep[e.g.,][]{2011SSRv..158..289G}. 
However, one complication is that our radial profile
    (see Equation~\eqref{eq_def_rhoEQ}) is not readily amenable to analytical treatment.
We therefore proceed with the numerical approach in 
    \citet{2021ApJ...908..230C}, computing the radial fundamental kink mode
    as a resistive eigenmode \citep[see][for conceptual clarifications]{2011SSRv..158..289G}.
The code outputs a period of $P_{\rm L}^{\rm EVP} = 57.3~{\rm s}$
    and a damping time of $\tau_{\rm L}^{\rm EVP} = 107~{\rm s}$,
    where the superscript indicates that these expectations are derived
    from an eigenvalue problem (EVP) perspective
    \footnote{The period of the radial fundamental kink mode
        reads $2L_0/c_{\rm k} \approx 55.8~{\rm s}$
        in the thin-tube (TT) thin-boundary (TB) limit, with $c_{\rm k}$
        being the kink speed \citep[e.g.,][]{2011SSRv..158..289G}. 
	This is not far from $P_{\rm L}^{\rm EVP}$. 
	We refrain from comparing $\tau_{\rm L}^{\rm EVP}$ with the TTTB expectation, 
	    because an expression is not available for our radial profile.
    The detailed formulation of a radial profile, however, 
        is known to impact the TTTB expectations
        for the damping time~\citep[e.g.,][]{2014ApJ...781..111S}.}. 
One sees that the expected period and damping time agree remarkably
    well with the best-fit values that we derive with the 
    3D time-dependent simulation (the left column of Figure~\ref{vx_parameter}).

\subsection{the Short-period Component}
\label{app_A2}
This subsection compares our short-period component
    with the EVP expectations for the first leaky kink mode.
We restrict ourselves to the piecewise constant version
    (i.e., $\alpha\to\infty$) of our equilibrium.
The pertinent flow field is emphasized, despite that 
    the expressions we offer are not new per se 
    \citep[e.g.,][]{1986SoPh..103..277C,2003SoPh..217...95C}. 
We work in a cylindrical coordinate system $(r, \theta, z)$, 
    denoting the equilibrium quantities with the subscript $0$. 
The interior and exterior are further discriminated by the subscripts~${\rm i}$ and ${\rm e}$.
There appears a set of primitive quantities $\{\rho_{\rm i, e}, p_{\rm i, e}, B_{\rm i, e}\}$,
    where $\rho$, $p$, and $B$ represent the mass density, thermal pressure, 
    and magnetic field strength, respectively. 
We  define the \Alf\ ($v_{\rm A}$), adiabatic sound ($c_{\rm s}$), 
    and tube speeds ($c_{\rm T}$) as
\begin{equation}
	\label{eq_def_CharSpeeds}
	v^2_{\rm Ai, e} = \dfrac{B^2_{\rm i, e}}{\mu_0 \rho_{\rm i, e}},\quad  
	c^2_{\rm si, e} = \dfrac{\gamma p_{\rm i, e}}{\rho_{\rm i, e}}, \quad   
	c^2_{\rm Ti, e} = \dfrac{v^2_{\rm Ai, e} c^2_{\rm si, e}}{v^2_{\rm Ai, e}+c^2_{\rm si, e}}, 
\end{equation}         
    with $\mu_0$ the magnetic permeability of free space 
    and $\gamma=5/3$ the ratio of specific heats.

Our EVP expectations are as follows. 
With kink motions in mind, we write any small-amplitude perturbation $\delta g$ as
\begin{equation}
	\delta g(r, \theta, z; t) 
	= \Re\left\{\tilde{g}(r) \exp\left[-i(\omega t-kz-\theta)\right]\right\},	
\end{equation} 
where $k$ denotes the real-valued axial wavenumber, and 
$\omega = \omega_{\rm R} + i \omega_{\rm I}$ represents the angular frequency.
Only temporally non-growing solutions are sought ($\omega_{\rm I}\le 0$). 
Defining 
\begin{equation}
	\label{eq_def_mu}	
	\mu_{\rm i, e}^2 
	= \frac{(\omega^2  - k^2 v_{\rm Ai, e}^2)(\omega^2 - k^2 c_{\rm si, e}^2)}
	{(v_{\rm Ai, e}^2+c_{\rm si, e}^2)(\omega^2 - k^2c_{\rm Ti, e}^2)},
\end{equation}
we further take $\omega_{\rm R} >0$ and $-\pi/2 < \arg\mu_{\rm i, e} \le \pi/2$.
A dispersion relation (DR) then writes
\begin{equation}
	\label{eq_DR}	
	\dfrac{\mu_{\rm i} R}{\omega^2-k^2 v^2_{\rm Ai}}
	\dfrac{J_1'(\mu_{\rm i} R)}{J_1(\mu_{\rm i} R)}
	= \dfrac{\rho_{\rm i}}{\rho_{\rm e}}	
	\dfrac{\mu_{\rm e} R}{\omega^2-k^2 v^2_{\rm Ae}}
	\dfrac{(H_1^{(1)})'(\mu_{\rm e} R)}{H_1^{(1)}(\mu_{\rm e} R)},
\end{equation}
	which accounts for both the trapped and leaky regimes. 
Here $J_n$ ($H_n^{(1)}$) denotes the Bessel (Hankel) function 
	of the first kind (with $n=1$).
The prime $'$ represents, say, 
	${\rm d}J_1(\mathfrak{z})/{\rm d}\mathfrak{z}$ with $\mathfrak{z}$
	evaluated at $\mu_{\rm i} R$.
Standard procedure further yields that
\begin{equation}
	\label{eq_tilde_pT}
	\tilde{p}_{\rm T}(r)= \dfrac{B^2_{\rm i}}{\mu_0} \times 
	\left\{ \begin{array}{ll}
		J_1(\mu_{\rm i}r)/J_1(\mu_{\rm i}R), 		  		& 0 \le r \leq R, \\[0.2cm]
		H^{(1)}_1(\mu_{\rm e}r)/H^{(1)}_1(\mu_{\rm e}R),    & r \ge R,
	\end{array}
	\right.
\end{equation}
where the arbitrarily scaled $\tilde{p}_{\rm T}$ denotes the Fourier amplitude of
the total pressure perturbation.
For both the interior and exterior, the Fourier amplitudes
for the radial and azimuthal speeds write
\begin{eqnarray}
	\tilde{v}_r(r) 
	&=& \dfrac{-i\omega}{\rho_0(\omega^2-k^2 v^2_{\rm A})}
	\dfrac{{\rm d}\tilde{p}_T}{{\rm d}r}, 
	\label{eq_tilde_vr} \\
	\tilde{v}_\theta(r) 
	&=& \dfrac{\omega}{\rho_0(\omega^2-k^2 v^2_{\rm A})}
	\dfrac{\tilde{p}_T}{r}. 
	\label{eq_tilde_vtht} 
\end{eqnarray}
Our EVP expectations amount to time-dependent perturbations
that are standing in both axial and azimuthal directions, the net
results being
\begin{eqnarray}
	v_r(r, \theta, z; t) 
	&=& A \sin(kz) \cos\theta 
	\Re\left[i \tilde{v}_r(r)    {\rm e}^{-i\omega t}\right]
	=
	\left[A \sin(kz) {\rm e}^{\omega_{\rm I} t}\right]     
	\left\{\cos\theta 
	\Re\left[i \tilde{v}_r(r)    {\rm e}^{-i\omega_{\rm R} t}\right]\right\},
	\label{eq_vr_tdep} \\
	v_\theta(r, \theta, z; t) 
	&=& A \sin(kz) \sin\theta 
	\Re\left[-\tilde{v}_\theta(r) {\rm e}^{-i\omega t}\right]
	=   
	\left[A \sin(kz) {\rm e}^{\omega_{\rm I} t}\right]     
	\left\{\sin\theta 
	\Re\left[-\tilde{v}_\theta(r) {\rm e}^{-i\omega_{\rm R} t}\right]\right\}.
	\label{eq_vtht_tdep}
\end{eqnarray}
Here $A$ is some arbitrary, real-valued, dimensionless magnitude. 

\begin{figure}    %%%%%%%%%%%%%%%%%% FIGURE 7
	\begin{center}
		\includegraphics[width=0.8\textwidth,clip=]{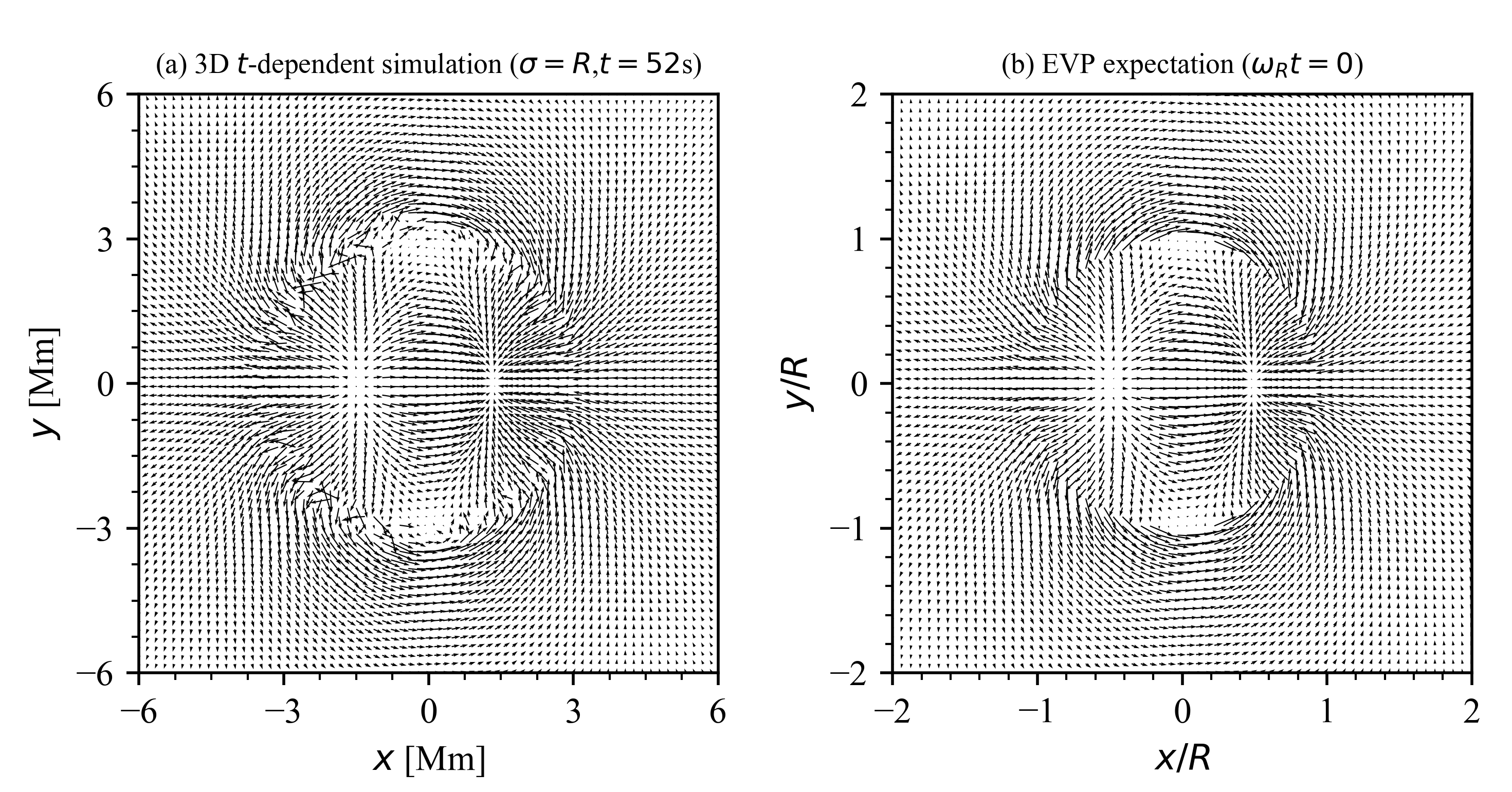}
		\caption{Velocity fields in the apex plane of 
			(a) the simulated short-period component at some representative instant 
			and (b) the eigenvalue problem expectation for the first leaky kink mode.}
		\label{v_field_compare}
	\end{center}
\end{figure}

We now compare our short-period component with the EVP expectations.
To start, plugging our equilibrium quantities into Equation~\eqref{eq_DR}
    yields a period of $P_{\rm S}^{\rm EVP} =5.96~{\rm s}$
    and a damping time of $\tau_{\rm S}^{\rm EVP} =150~{\rm s}$.
This pertains to the first leaky mode, namely the one that possesses 
    the lowest $\omega_{\rm R}$ among all modes with $\omega_{\rm I}<0$
    \footnote{Some analytical expressions are available that 
    approximately solve Equation~\eqref{eq_DR} in some appropriate limits
    \citep[e.g.,][]{2003SoPh..217...95C,1982SoPh...75....3S}.
	We choose to solve Equation~\eqref{eq_DR} exactly, because
       the eigenfunctions are also of interest.}.
The expected period is in good agreement with the best-fit values printed
    in the left column of Figure~\ref{vx_parameter}.
The best-fit damping times, on the other hand, are somehow shorter than expected,
    which is intuitively understandable because a radially discontinuous cylinder
    leads to more efficient wave trapping. 
Figure~\ref{v_field_compare}a further displays the velocity fields 
    in the apex plane of our short-period component at some representative instant.
Plotted in Figure~\ref{v_field_compare}b is the pertinent EVP expectation, 
    namely the flow field constructed with the terms in the braces    
    of Equations~\eqref{eq_vr_tdep} and~\eqref{eq_vtht_tdep} at $\omega_{\rm R} t = 0$.
The two sets of flow fields are remarkably similar.
Overall, we conclude that the short-period component can be confidently identified
    as the first leaky kink mode. 

\section{Effect of Varying $\sigma$}
\label{sec_AppB}
\begin{figure}    %%%%%%%%%%%%%%%%%% FIGURE 6
	\begin{center}
		\includegraphics[width=0.32\textwidth,clip=]{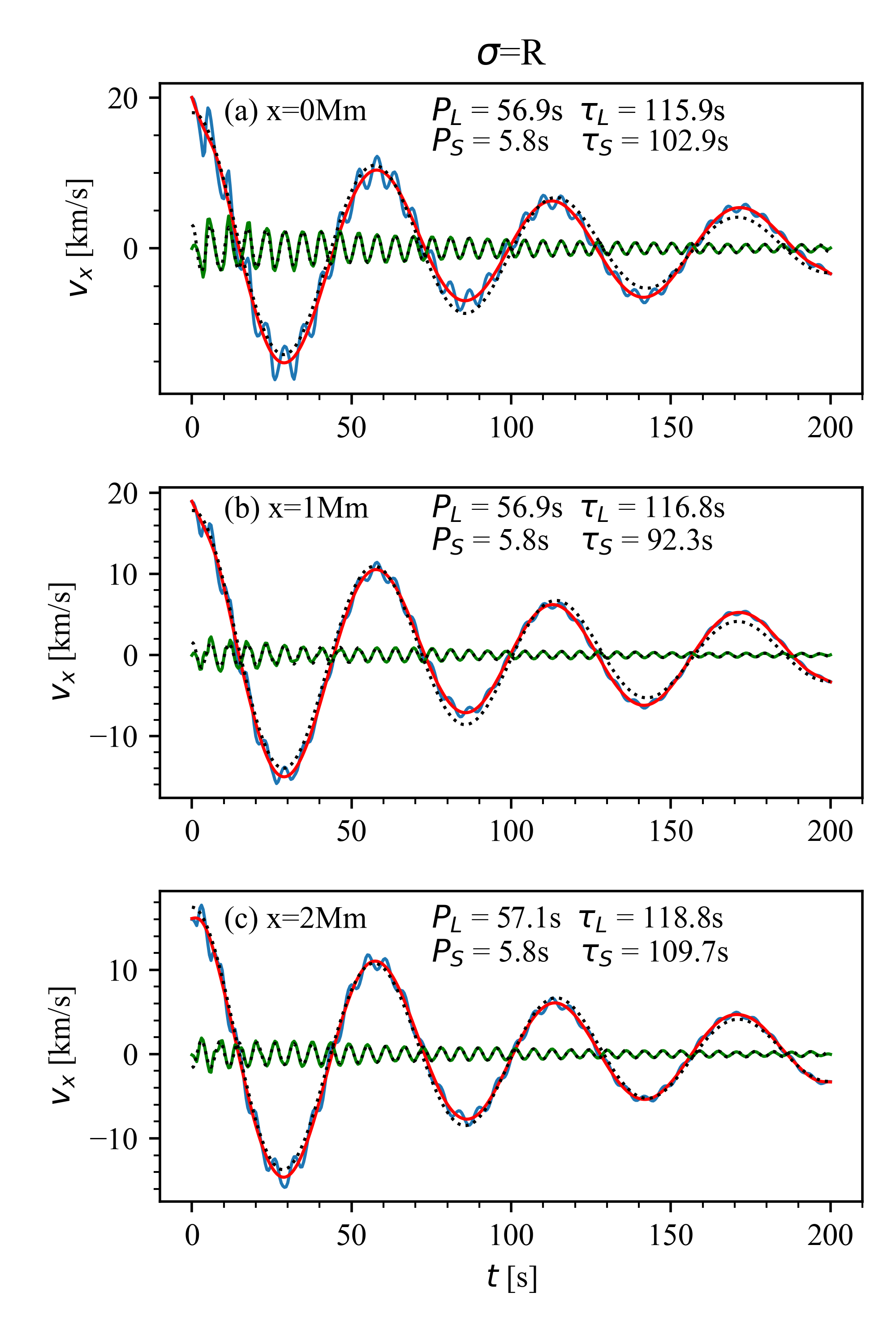}
		\includegraphics[width=0.32\textwidth,clip=]{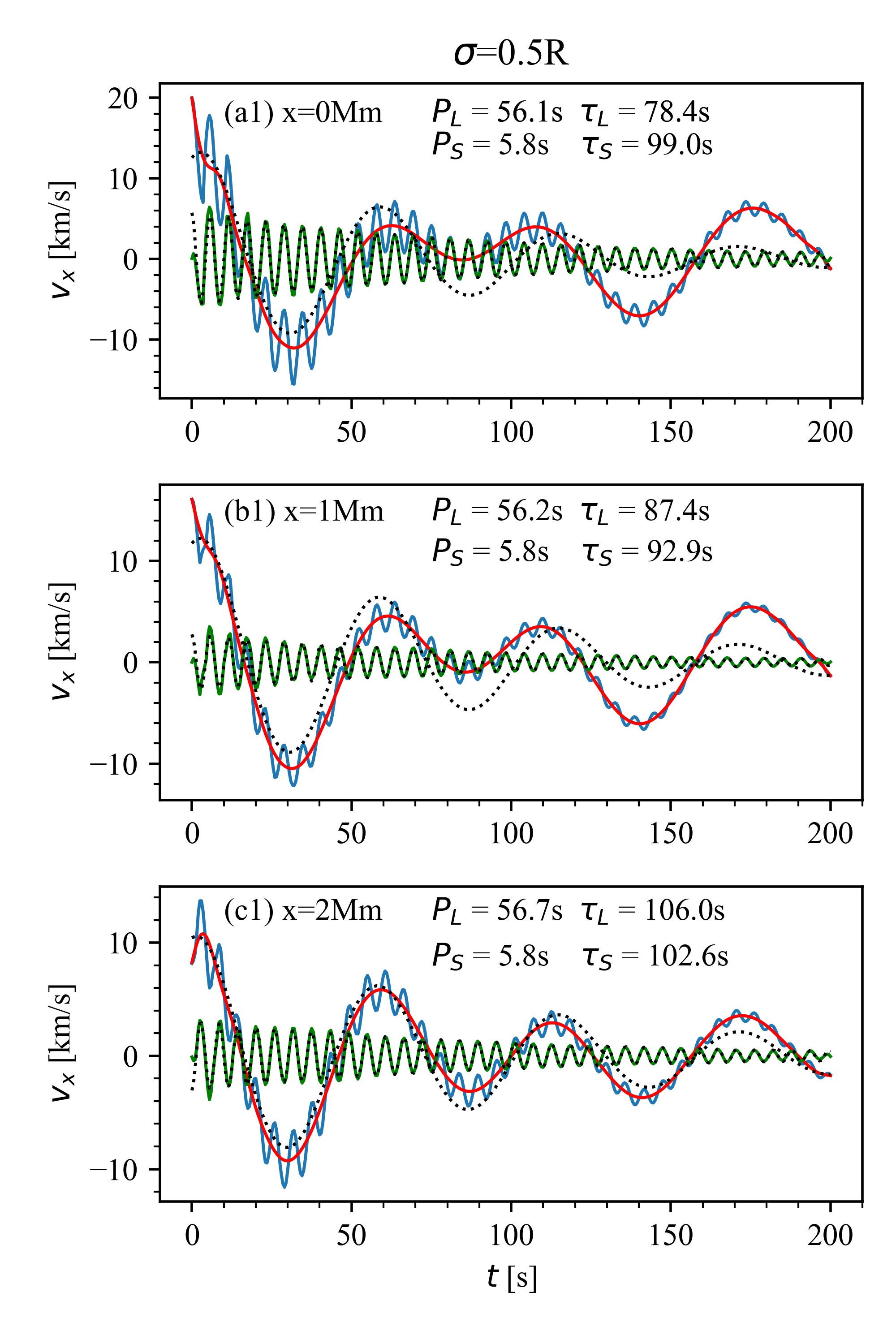}
		\includegraphics[width=0.32\textwidth,clip=]{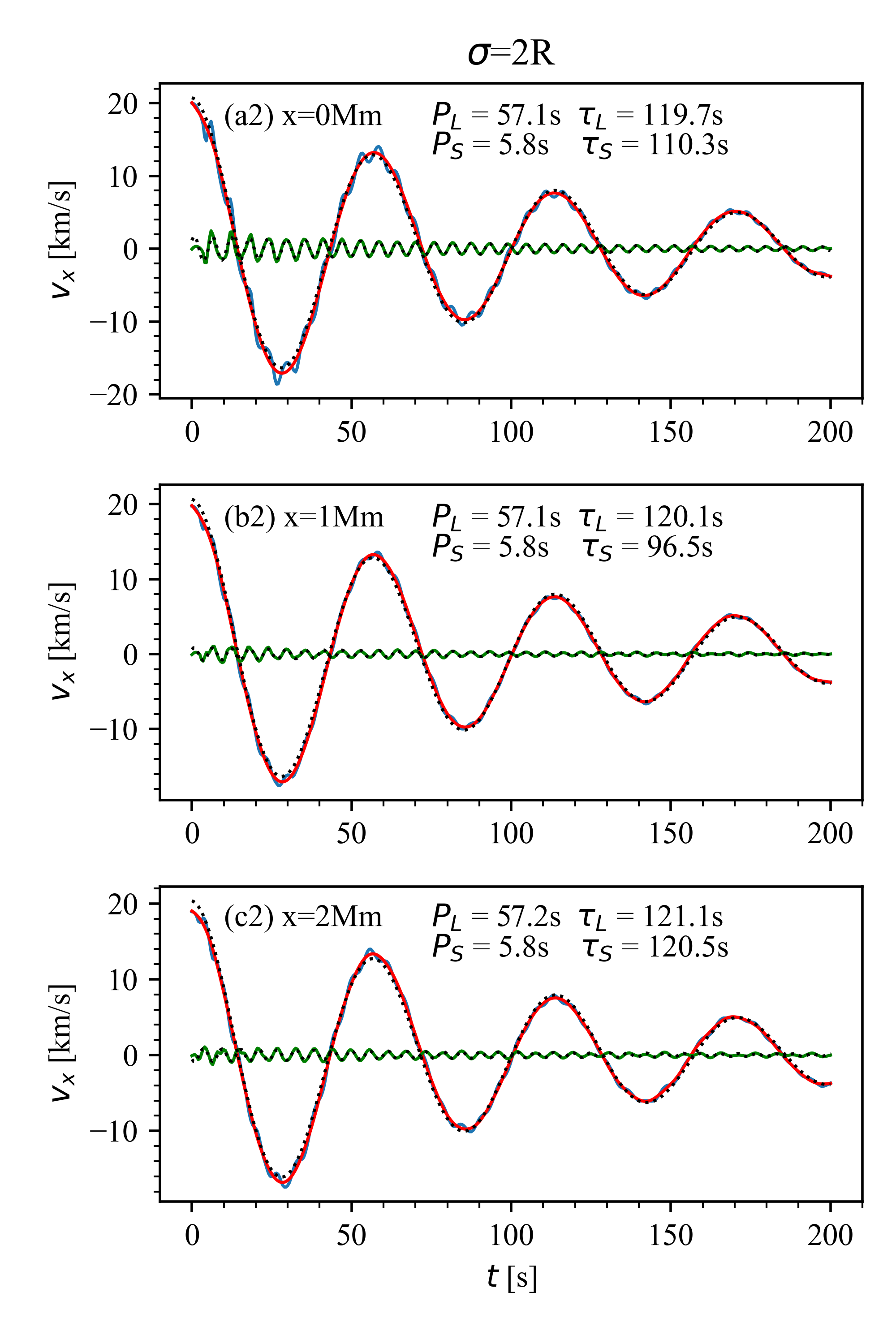}
		\caption{Left column: Temporal evolution of $v_x$ sampled in the apex plane at (a) $(x,y)=(0,0){\rm Mm}$, (b) $(x,y)=(1,0){\rm Mm}$, and (c) $(x,y)=(2,0){\rm Mm}$ for the case $\sigma=R$. Any blue curve represents the original signal, while the red and green curves give the low-pass and high-pass components, respectively.
		The black dotted lines display the fitting curves of the high-(low-) pass components
		    by an exponentially damping sinusoid, with the best-fit periods and damping times shown in each panel.
		Middle column: same as the left but for $\sigma=0.5R$.
		Right column: same as the left but for $\sigma=2R$.
		}
		\label{vx_parameter}
	\end{center}
\end{figure}

This section examines how the relative importance 
    of the long- and short-period components varies
    when we vary the spatial extent of
    the initial perturbation (i.e., $\sigma$ in Equation~\ref{eq2}). 
Two additional simulations are performed, 
    one with $\sigma=0.5R$ and the other with $\sigma=2R$. 
These additional computations are shown in the middle and right columns of
    Figure~\ref{vx_parameter} in the same format as our reference
    results (with $\sigma = R$). 
Two pronounced periodicites can be readily told apart in 
    any sampled $v_x$. 
The following features stand out for the short-period components.
Firstly, their best-fit periods and damping times 
     vary little from one case to another, indicating the robust excitation
     of the first leaky kink mode. 
Secondly, the short-period component tends to be stronger for
     smaller values of $\sigma$, suggesting 
     that leaky modes can receive a larger fraction of the energy imparted by the initial perturbation when it is more localized.      

Some subtlety exists for the long-period component when $\sigma=0.5R$.
Overall, the temporal behavior for $v_x$ deviates from
     an exponentially damping sinusoid, in contrast to what happens for other values of $\sigma$
     that we examine.
We refrain from explaining why at this time of writing.
Rather, we note that this deviation is not a numerical artifact but persists even when 
     we adopt a considerably larger computational domain. 

%\section{Effect of the simulation domain}




\bibliographystyle{apj}
\bibliography{export-bibtex}

%\begin{thebibliography}{}
	
%\end{thebibliography}


\clearpage

\clearpage

\clearpage

\clearpage

\clearpage

\clearpage
\end{document}